\documentstyle[12pt]{article}

\begin{document}
\def\la{\mathrel{\mathpalette\fun <}}
\def\ga{\mathrel{\mathpalette\fun >}}
\def\fun#1#2{\lower3.6pt\vbox{\baselineskip0pt\lineskip.9pt
        \ialign{$\mathsurround=0pt#1\hfill##\hfil$\crcr#2\crcr\sim\crcr}}}
\newcommand {\eegg}{e^+e^-\gamma\gamma~+\not \! \!{E}_T}
\newcommand {\mumugg}{\mu^+\mu^-\gamma\gamma~+\not \! \!{E}_T}
\renewcommand{\thefootnote}{\fnsymbol{footnote}}
\bibliographystyle{unsrt}

\begin{flushright}
UMD-PP-99-014\\
\end{flushright} 
\begin{center}
{\Large \bf Limits on Pauli Principle Violation by Nucleons}

\vskip1.0cm

{\bf E. Baron$^1$, Rabindra N. Mohapatra$^2$, and Vigdor L. Teplitz$^3$}

\vskip1.0cm

{\it$^{(1)}${ Department of Physics and Astronomy, University of
Oklahoma, Norman, 
OK-73019, USA.}}

{\it$^{(2)}${ Department of Physics, University of Maryland,
College Park, MD-20742, USA.}}

{\it $^{(3)}${Department of Physics, Southern Methodist University,
Dallas, TX-75275, USA}}

\end{center}

\begin{abstract}

We consider nuclei produced in core collapse supernovae and subjected
 to a high neutron flux.  We show that an accelerator mass
 spectrometry experiment that searched for traces of anomalous iron
 isotopes could set limits on the order of $10^{-20}-10^{-25}$ on (or
 perhaps discover) Pauli principle violation by neutrons.  A similar
 search for anomalous Co isotopes could set limits in the range
 $10^{-13}-10^{-18}$ on Pauli principle violation by protons.  We show
 that existing data on Oxygen can be used to set a limit of about
 $10^{-17}$ in one proposed model of such violation.
\end{abstract}

\newpage
\renewcommand{\thefootnote}{\arabic{footnote})}
\setcounter{footnote}{0}
\addtocounter{page}{-1}
\baselineskip=24pt

 \vskip0.5cm

	The Pauli exclusion principle, one of the the fundamental
principles on which our basic understanding of the inner workings of
systems ranging from atoms and nuclei to solids and liquids all the
way to the stars and the universe rests, is clearly valid to a high
degree of precision. Two simple examples in the domain of chemistry
and physics are the patterns observed in the periodic table and the
success of the shell model for nuclei. Until recently, however, there
were hardly any quantitative tests of this fundamental principle. The
best limit on possible deviation from the Pauli principle before 1987
was from an early experiment of Goldhaber and Scharff-Goldhaber
\cite{gold} in which beta rays from C$^{14}$ were allowed to impinge
on lead and a search was made for K-shell x-rays. This bound on
possible violation of the Pauli principle (characterized by a parameter
$\beta^2$ in this article and earlier \cite{gm}) by electrons was at
the level of 3\%. In 1968, Fishbach, Kirsten and Schaffer
\cite{fishbach} conducted a search for anomalous $^9Be$ and put an
upper limit on the atmospheric density of such elements. Their result
could be interpreted \cite{gm} to derive an upper limit of
$\beta^2\leq 2\times 10^{-3}$.

 In 1987-90, several theoretical papers postulated 
ways to incorporate small deviations from Bose and Fermi statistics into 
quantum mechanics and field theories \cite{ik,gm,rnm,gm2}. At this time, it 
was noted \cite{gm} that atomic  
spectroscopy could improve this bound for electrons to the level of $10^{-7}$.
Inspired by the theoretical interest in the subject, 
several dedicated experiments were planned around 1987-88 and have now 
considerably improved the limits 
for electrons \cite{kell,snow,poma}. These limits are now at the level of
$10^{-26}$ - $10^{-27}$. As far as other particles go, strong limits for 
protons (though much weaker than for electrons) have been extracted by 
Plaga \cite{plaga} from considerations of energy generation in the Sun. We 
have listed these limits in the first four rows of Table 1. Note that
to date no limit has been established for neutrons. There also exist 
limits on deviations from bose statistics 
for bosonic systems such as pions \cite{gm}, photons \cite{bos} and 
spin zero atoms \cite{bos2}, which we do not address here. 

Meanwhile, on the theoretical front it was shown that deviation from Fermi 
and Bose statistics can be embodied in a satisfactory manner \cite{gm2}
using the following ``quonic'' commutation relations \cite{rnm} among the 
annihilation and creation operators ($a_i, a^{\dagger}_i$):
\begin{eqnarray}
a_ia^{\dagger}_j-q a^{\dagger}_ja_i = \delta_{ij}
\end{eqnarray}
The Hilbert space of states for this model is positive definite, though 
it leads to a non-local field theory \cite{gm2} and a number operator
which is non-polynomial. Also we note that for the fermionic particles,
$q=-1+\beta^2$.

 In view of the fundamental role of the Pauli principle, it appears to us of 
considerable interest to expand the domain of our knowledge of the limits on
Pauli violation (PV for short) to as many particles as possible.

The purpose of this note is to point out  potential, high precision
tests of the Pauli principle, for neutrons and protons, making use of the 
high neutron and proton fluxes in a type II supernova and also to derive a 
bound from already existing limits on anomalous oxygen isotope by Hemmick et 
al. \cite{hemmick}. At the risk of removing all suspense, we have included 
our results from this paper in Table 1.

\vbox{
\begin{center}
			Table 1.  Pauli Principle Tests
\begin{tabular}{llll}
\hline
Particle& Reference& Method& Upper limit on $\beta^2$\\
\hline
Electron & Goldhaber and Goldhaber \cite{gold} & K-capture & 0.03 \\
Electron & Fishbach et al \cite{fishbach} & Anomalous $^9Be$ & $2\times 
10^{-3}$\\
Electron& Greenberg and Mohapatra \cite{gm} & Spectroscopy&  $10^{-7}$\\
Electron& Kelleher \cite{kell} & Spectroscopy & $10^{-7}$\\
Electron& Ramberg and Snow \cite{snow} & Anomalous X-rays& $10^{-27}$\\
Electrons & Pomansky et al. \cite{poma} & Mass spectrometry& $10^{-26}$\\
Proton  &  Plaga \cite{plaga} & Solar p-p cycle & $10^{-15}$\\
Neutron& This paper& Supernova Oxygen \& &$10^{-17}$\\
&   & Mass spectrometry& (model-dependent)\\
Neutron& This paper& Supernova Iron \&     & $ 10^{-25}$\\
       &                   &      Mass spectrometry&       is possible\\

Proton& This paper& Supernova Cobalt \&    & $10^{-18}$\\
       &                   &      Mass spectrometry&       is possible\\

\hline
\end{tabular}
\end{center}
}

Our argument is as follows. About half of the terrestrial iron is
produced during core collapse of massive stars to form Type II (or
Type Ib/c) supernovae \cite{thiel}.  There, nickel is produced and
there is also a significant flux of free neutrons.  After the
supernova bounce, the ensuing shock wave ejects the nickel which
decays to cobalt which, in turn, decays to iron.  In the supernova
explosion there is approximately a fraction of a second during which
the nickel is formed and exposed to a high flux of neutrons with
thermal energies of about $0.5$ MeV. The exact value of the neutron
flux depends on the electron fraction of the material ejected $Y_e$,
which in turn depends on the uncertain details of the supernova
process. The other half of the terrestrial iron is produced in Type Ia
supernovae by the same process, but the density is lower and the free
neutron density is significantly lower since the supernova occurs as a
thermonuclear explosion of a C+O white dwarf and there is little time
for weak interactions to occur. Thus we will concentrate on iron
produced in core collapse supernovae. During the explosion, Pauli
violation (PV) could permit the process
\begin{equation}
 n+^{A-1}Ni\rightarrow^{A}Ni^*+\gamma,  \label{nickel}
\end{equation}
where $^ANi^*$ contains an extra neutron in an already occupied shell model
state.  For example it could have 3 neutrons in the $1-s$ shell.  Assuming that
this (inner) configuration was not disturbed in the electron capture
transitions from $Ni^*$ to $Fe^*$ (and did not inhibit those
transitions), we would have today about the same relative 
abundance, up to a factor of two, of anomalous iron. Normal stable iron 
isotopes have A= 54, 56, 58. The unstable isotopes with A=55 and 59 are 
the products of shortlived isotopes and also have  very short halflives. 
Hence they are not expected to occur in terrestrial radioactivities. Thus 
experimental detection of these two, which is well within the capabilities of
accelerator mass spectrometry, would be an indication of Pauli violation.
The abundances of $^{54}Fe$ and $^{58}Fe$, from which $^{55}Fe^*$ and 
$^{59}Fe^*$ would be made, effectively, are 5.9\% and 0.33\%.

We turn now to the question of what abundance ratio, $Fe^*/Fe$, we can
expect for a given amount of Pauli violation (PV).
  The probability that a
neutron will be captured into a PV state is given by
\begin{equation} 
 P_V=\beta^2\sigma n v t\label{PV}
\end{equation}
where $\beta^2$ measures the suppression of the cross section $\sigma$ that
would obtain in the absence of the Pauli principle, $n$ is the neutron
density,  $v$ is the velocity of the neutrons, and $t$ is the time available.

 There should be three contributions to the cross section for the
process of Eq(\ref{nickel}): magnetic dipole radiation from neutron
spin flip; magnetic dipole radiation from Ni spin flip; and electric
dipole radiation from Ni.  Blatt and Weisskopf \cite{bw} give a useful
discussion of capture and disintegration cross sections which we
follow.  We keep only the first contribution (neutron spin flip) since
the others are smaller, and we attempt to be conservative in
estimating the rate.  We use the zero range approximation (Eq (4.27)
of \cite{bw}).  We ignore possible scattering enhancement factors,
i.e. $(1-a\gamma)^2$ where $a$ is the scattering length and
$\gamma=(2\mu B)^{1/2}$ with $\mu$ the n-Ni reduced mass, about the
nucleon mass, and $B$ the anomalous binding energy, on the order of
$20$~MeV.  The capture cross section is then
\begin{eqnarray}
\sigma&=&\pi(e^2/\hbar c)(\hbar
      c/M_nc^2)^2(B/M_nc^2)(2B/E_n)^{1/2}\label{sigcap}\\ 
      &=&10^{-30}{\rm cm}^2,\nonumber
\end{eqnarray}
where $E_n=0.5$ MeV is the neutron energy.
For the density of neutrons, we have attempted to be 
conservative in our estimate,  and are guided by the results from 
assuming  nuclear  
statistical equilibrium (NSE) (see Table 2 for some typical values in the 
region of the core that is ejected by the shock wave based on the model 
of Cooperstein, Kahana and Baron\cite{coop}).

In the supernovae the material is not ejected with NSE abundances, but
rather undergoes an alpha rich freeze out that has been suggested to
be in 
quasi-statistical equilibrium \cite{woos-meyer} (more alpha particles at a 
given ($\rho$,T,$Y_e$) than would be predicted by NSE). The neutron and proton
densities drop precipitiously with temperature as the material
expands. While the  neutron flux is not determined by NSE, it is
indicative. As the temperature drops during expansion, NSE is
an increasingly poor assumption and so one has to both use a QSE code and
know the expansion, T, and Ye history of the material, which is totally
uncertain. Therefore the  number  $X_n$ used by us is simply a guess
``guided'' by  
the results in Table 2 which is based on  assuming a reasonable equation of 
state for supernova material. We take $\rho\sim10^9$~g~cm$^{-3}$, and a 
neutron mass fraction $X_n\sim 10^{-4}$, giving
$n\sim10^{29}$~cm$^{-3}$.  The thermal velocity of $0.5$~MeV neutrons is about
 $10^9$~cm~s$^{-1}$ 
and the Ni is exposed to neutrons for about a 0.1~second.  This gives
\begin{equation}
 P_V\sim10^{8}\beta^2.\label{pvno}
\end{equation}
That is, we have
\begin{equation}
\beta^2<10^{-8}{\rm EB}\label{pvbnd}
\end{equation}
where EB stands for an experimental upper bound on the relative abundance of
anomalous iron. The above estimate is for the dominant isotope of iron 
i.e. $^{56}Fe$. Since we are interested in $^{59}Fe^*$, we must multiply 
this probability by the factor $3\times 10^{-3}$ to take into account the
low abundance of $^{58}Fe$ (which presumably implies the corresponding 
low abundance of the parent $Ni$ isotope which participated in the PV 
reaction).  Based on experience with such bounds for other elements
\cite{hemmick}, we can expect that limits in the range $10^{-13}$ to
$10^{-18}$ are readily achievable.  Such an experiment would then set a PV
limit for $\beta^2$ on the order of $3\times 10^{-19}$ to $3\times 10^{-24}$.
If we use a similar argument for anomalous $^{55}Fe^*$, we would consider 
a parent $Fe$ nucleus to be $^{54}Fe$ whose abundance is 5.9\%. Thus, looking
for $^{55}Fe^*$ would provide bound of $1.5\times 10^{-20}$ to $1.5\times 
10^{-25}$.
 \begin{center}
Table 2. Illustrative Abundances for Supernova Conditions
\begin{tabular}{llllllll}
\hline
$\rho$ (g cm$^{-3})$&T (MeV)& $Y_e$ & $X_n$ &$X_p$&$X_\alpha$&$X_{\rm
Fe}$ &$X_{\rm Ni}$\\
\hline
$10^{10}$& 1.0& 0.49&5.2 (-3)& 1.8 (-2)& 5.2 (-1)& 1.8 (-1)&5.0 (-4)\\
$10^{10}$&1.0&0.495&4.4 (-3)&2.1 (-2)&5.5 (-1)&1.6 (-1)&9.5 (-4)\\
$10^{10}$&1.0&0.50&3.7 (-3)&2.5 (-2)&5.7 (-1)&1.4 (-1)&1.6 (-3)\\
$10^9$&1.0&0.49&7.2 (-2)&5.2 (-2)&8.8 (-1)&5.6 (-10)&7.3 (-14)\\
$10^9$&1.0&0.495&6.6 (-2)&5.6 (-2)&8.8 (-1)&4.2 (-10)&7.4 (-14)\\
$10^9$&0.5&0.50&1.1 (-8)&4.2 (-3)&5.4 (-3)&8.0 (-5)&7.7 (-1)\\
$10^8$&1.0&0.49&3.0 (-1)&2.8 (-1)&4.3 (-1)&1.4 (-27)&3.0 (-31)\\
$10^8$&1.0&0.495&2.9 (-1)&2.8 (-1)&4.3 (-1)&1.3 (-27)&3.1 (-31)\\
$10^8$&0.5&0.49&5.2 (-7)&7.3 (-3)&4.1 (-2)&1.3 (-2)&1.6 (-1)\\
$10^8$&0.5&0.495&3.5 (-7)&1.1 (-2)&4.2 (-2)&4.8 (-3)&2.8 (-1)\\
$10^8$&0.3&0.50&0.0 &5.5 (-5)&5.3 (-6)&8.7 (-10)&9.9 (-1)\\
\hline
\end{tabular}
\end{center}
%In the above we have been somewhat simplistic and cavalier in assuming
%that the nickel is bathed in a large flux of free neutrons for a full
%second when the ejected nickel is in fact produced by the explosive
%burning of Si and the total neutron flux and exposure time depend on
%the exact value of $Y_e$. However, the Pauli violation can occur in
%any $(n,\gamma)$ reaction in the complicated process of explosive
%silicon burning. To take into account these uncertainties we reduce
%the upper bound by a factor of $10^6$. 

We need to emphasize the fact that the estimate of neutron density in the 
relevant region is subject to considerable uncertainty. There is as yet 
no successful model of supernova explosion. The assumption of NSE 
is approximate, but should not be unreasonable given that  silicon burning
to Ni is explosive and takes place at temperatures $T >
0.5$~MeV. In reality we would need to consider all $(n,\gamma)$
reactions that occur in the nucleosynthetic process and we have tried
to be conservative in our assumptions in order to derive a meaningful
limit. The time  
available in the neutron bath and the density trajectory of the
ejected material is a complicated function of explosion details.
We believe however that these effects should not change the result quoted by 
more than a factor of 100. It is important to note that, even for the worst
case scenario, we can expect a meaningful bound on $\beta^2$ for neutrons 
from these considerations.

 Table 1 compares this result with other PV limits.  The closest
comparison is to the work of Plaga \cite{plaga} who considered the effect 
of a
possible triplet s-wave (hence PV) proton-proton bound state on the solar p-p
cycle, achieving a bound of $10^{-15}$ from the (assumed) approximate validity
of the standard solar model.

	In addition to the large neutron flux, the supernova has a large proton
flux.  Indeed, as noted above the proton flux is sufficiently large to overcome
the coulomb barrier and convert most of the iron to nickel.  This tells us
that, for protons, the coefficient of $\beta^2$ in Equation (3) must be on the
order of one (or larger). As in the neutron case, we are assuming that
the extra 1-s proton neither inhibits the decays nor is disturbed by
them. For the PV case, because the proton will be much
more deeply bound, we might expect it to be considerably larger.  Here we just
assume it is of order $1$.  We thus see immediately that the bound on $\beta^2$
is just equal to the bound on the abundance of the anomalous isotope, Co*.  One
could argue a $10^3$ enhancement in view of the fact that the abundance of Co*
will be a fraction of the abundance of Fe while the abundance of normal Co is,
at least on Earth, $10^{-3}$ that of Fe. There are no long lived Co isotopes
besides the stable $^{59}Co$. The other isotopes are 
products of shortlived isotopes and also have very short halflives. Moreover 
the decay energies of Co 
isotopes are much less than the 20 MeV or more that should be gained by
putting the final proton in the $1s$ state of Cobalt. Thus we expect that 
any detection of Z=27 nucleus with $A\neq 59$ in a non-irradiated target
would indicate $Co^*$ and hence Pauli violation. 

 Finally, we note that existing data can be used to set a limit of
interest for one PV model.  One of us (RNM) has 
 shown that if Pauli violation is characterized by the 
commutation relations (see \cite{rnm}) in Eq. (1),
then a matrix representation for the annihilation operators is given by 
 \begin{equation}
 a = \left(\begin{array}{cccc}
            0 & 0 & 0 & ...\\
     1 & 0 & 0 &..\\
     0 &\beta_1& 0 &..\\
     0 & 0 & \beta_2 &..\\
     . & . & . & . \\
     . & . & . & . \\
    \end{array}\right)\label{mat}
\end{equation}
with $\beta_n^2=1+q+... +q^n$.  For $q$ close to $-1$, one sees that for odd
$n$, $\beta^n$ is of the order $\beta_1^{(n+1)/2}$ while, for even $n$,
$\beta_n$ is 
of order $1$.  Thus, in a neutron rich environment, the second neutron is added
relatively easily.  The presupernova star has an oxygen shell with a neutron
abundance around $n\sim10^{20}$.  Eq (\ref{PV}) gives a capture probability of
about $0.1\beta^2$.  Middleton et al, some time ago, put a mass spectrometer 
limit on on anomalous oxygen, down to 20 AMU of about $O^*/O<10^{-18}$ which
gives a PV bound, within the model of  \cite{rnm}, of $\beta_1^2<10^{-17}$.
We note however that the above matrix representation was derived in \cite{rnm}
for the case of a single oscillator. It is therefore not clear whether it
will remain valid in an actual field theory generalization. One possible 
argument which may suggest its validity in a general field theory, can be 
given as follows.  If we put the quon fields in 
a thermal bath, in the low temperature limit, we expect the partition 
function for the system to be dominated by the lowest energy level which 
will then be described by a single oscillator. For this case then the 
single oscillator representation should remain valid.

 In summary, we suggest use of mass spectrometry to limit Pauli
principle violations for neutrons and protons by exploiting the high 
nucleon fluxes in supernovae.  A neutron limit of interest on a specific 
model is derived from existing data on Oxygen isotopes and  much lower, 
model-independent limits from new experiments with iron and cobalt are
possible. 

ACKNOWLEDGMENTS

 We have benefited greatly from conversations with J. Cowan, D. Elmore, E. 
Fischbach, and O.W. Greenberg.  The work of R. N. M. is supported by the
National Science foundation under grant PHY-9802551 and that of V. L. T. by
DOE under grant DE-FG03-95ER40908. The work of E. B. is supported in
part by NSF grants AST-9417242, AST-9731450; NASA
grant NAG 5-3505, and an IBM SUR grant.


\begin{thebibliography}{99}

\bibitem{gold} M. Goldhaber and G. Scharff-Goldhaber, Phys. Rev. {\bf 
73}, 1472 (1948).

\bibitem{gm} O. W. Greenberg and R. N. Mohapatra, Phys. Rev. {\bf D 39}, 2032
(1989).

\bibitem{fishbach} E. Fishbach, T. Kirsten and O. Shaeffer, Phys. Rev. Lett.
{\bf 20}, 1012 (1968).

\bibitem{ik} A. Yu Ignatiev and V. Kuzmin, Sov. Journ. Nucl. Physics, 
{\bf 46}, 444 (1987); O. W. Greenberg and R. N. Mohapatra, Phys. Rev. 
Lett. {\bf 59}, 2507 (1987); {\it ibid} {\bf 62}, 712 (1989); L. Okun, 
JETP Lett. {\bf 46}, 529 (1987); A. V. Gorkov, Phys. Lett. {\bf A 137}, 7 
(1989).

\bibitem{rnm} O. W. Greenberg, Phys. Rev. Lett. {\bf 64}, 705 (1990); R. 
N. Mohapatra, Phys. Lett. {\bf 242}, 407 (1990); D. I. Fivel, Phys. Rev. 
Lett. {\bf 65}, 3361 (1990).

\bibitem{gm2} O. W. Greenberg, Phys. Rev. {\bf D 43}, 4111 (1991); 
Physica {\bf A 180}, 419 (1992); O. W. Greenberg and R. Hilborn, 
hep-th/9808106. 

\bibitem{kell} K. Deilamian, J. D. Gillaspy and D. Kelleher, Phys. Rev. 
Lett. {\bf 74}, 4787 (1995); K. Deilamian, Ph. D. thesis, Univ. of Wisconsin,
Madison, 1991.

\bibitem{snow} E. Ramberg and G. Snow, Phys. Lett. {\bf B 238}, 438 (1990).

\bibitem{poma} E. Nolte et al., Technical Univ. of Munich preprint (1989).

\bibitem{plaga} R. Plaga, Z. Phys. {\bf A 333}, 397 (1990).

\bibitem{bos} A. Ignatiev, G. C. Joshi and M. Matsuda, UM-P-94/62;

\bibitem{bos2} R. Hilborn and Candice L. Yuca, Phys. Rev. Lett. {\bf 76},
2844 (1996); M. d Angelis, G. Gagliardi and L. Gianfrani, Phys. Rev. Lett.
{\bf 76}, 2840 (1996).

\bibitem{hemmick} T. K. Hemmick et al., Phys. Rev. {\bf D 41}, 2074 (1990).
R. Middleton et al. Phys. Rev. Lett. {\bf 43}, 429 (1979); R. Middleton 
et al. ANL-PHY-81-1 (1981).

\bibitem{thiel} F.-K. Thielemann and K. Nomoto and M. Hashimoto, 1998, 
ApJ, 460, 408;
F. Timmes and S. Woosley and T. Weaver, 1995, ApJS, 98, 617;
T. Tsujimoto and K. Nomoto and Y. Yoshi and M. Hashimoto and
S. Yanagida and F.-K. Thilemann, 1995, MNRAS, 277, 945.

\bibitem{bw} J. Blatt and V. Weisskopf, {\it Theoretical Nuclear Physics},
Dover (1988).

\bibitem{coop} J. Cooperstein and E. Baron in {\it Supernovae},
A. Petschek, ed., Springer-Verlag, New York, 1990, p. 213; J. Cooperstein,
Nucl. Phys. {\bf A 438}, 722 (1985); 
  E. Baron and J. Cooperstein and S. Kahana, Nucl. Phys. {\bf A 440}, 744 
(1985); Phys. Rev. Lett. {\bf 55}, 126 (1985).   

\bibitem{woos-meyer} S. Woosley and R. Hoffman, ApJ {\bf 395}, 202 
(1992); B. Meyer, T. Krishnan, and D. Clayton, ApJ  {\bf 498}, 808 (1998).

\end{thebibliography}
\end{document}